%% file: main.tex
\documentclass[11pt,a4paper]{article}

\usepackage[T1]{fontenc}
\usepackage{lmodern}
\usepackage[margin=2.5cm]{geometry}
\usepackage[numbers,sort&compress]{natbib}
\usepackage{amsmath,amssymb,amsfonts}
\usepackage{graphicx}
\usepackage{textcomp}
\usepackage{url}
\usepackage{booktabs,multirow,makecell,array}
\usepackage[nolist]{acronym}
\usepackage{authblk}
\usepackage[hidelinks]{hyperref}
\hypersetup{
  pdftitle={Tissue-Aware Digital Twins for Fluence and Skin Tone Modeling in Photoacoustic Imaging},
  pdfauthor={Moritz Schillinger et al.}
}

\title{Tissue-Aware Digital Twins for Fluence and Skin Tone Modeling in Photoacoustic Imaging}

\author[1]{Moritz Schillinger\thanks{Corresponding author: \texttt{moritz.schillinger@fau.de}}}
\author[1]{Maja Schlereth}
\author[2]{Felix Wachter}
\author[3]{Jérémie Gillet}
\author[3]{Guillaume Zahnd}
\author[4]{Adrian Buehler}
\author[5]{Niklas Holzwarth}
\author[2]{Ferdinand Knieling}
\author[4]{Ulrich Rother}
\author[1,6]{Katharina Breininger}

\affil[1]{Department Artificial Intelligence in Biomedical Engineering (AIBE), Friedrich-Alexander-Universität Erlangen-Nürnberg, 91052 Erlangen, Germany}
\affil[2]{Department of Pediatrics and Adolescent Medicine, University Hospital of Erlangen, 91054 Erlangen, Germany}
\affil[3]{iThera Medical GmbH, 81379 Munich, Germany}
\affil[4]{Department of Vascular Surgery, University Hospital of Erlangen, 91054 Erlangen, Germany}
\affil[5]{Division of Intelligent Medical Systems (IMSY), German Cancer Research Center (DKFZ), 69120 Heidelberg, Germany}
\affil[6]{Center for AI and Data Science (CAIDAS), Julius-Maximilians-Universität Würzburg, 97070 Würzburg, Germany}

\date{}

\begin{document}

\maketitle

\begin{abstract}
Quantitative \ac{pai} remains challenging because light fluence decays with depth and varies with wavelength, introducing strong biases in measured signal intensities. These effects distort chromophore estimates, obscure deeper structures, and cause skin-type-dependent variability, which can compromise diagnostic reliability. To address these limitations, we present a subject-specific, physics-informed framework for fluence correction using anatomical information from co-registered ultrasound scans. A deep neural network segments key tissues to construct individualized digital twins with spatially varying optical parameter maps. Monte Carlo simulations generate \textit{in silico} fluence estimates, which are used to correct the reconstructed \textit{in vivo} photoacoustic image. Epidermal melanin content is estimated from the segmented epidermis and incorporated into the model to account for inter-subject variation in skin pigmentation. Evaluated on simulations and clinical datasets, our method reduces depth-related bias and skin-type-associated signal variation while maintaining or improving diagnostic performance. This enables large-scale, automated, and more equitable quantitative \ac{pai} analysis.
\end{abstract}


\input{body}

\section*{Disclosures}
J.G. and G.Z. were employees of iThera Medical GmbH (Munich, Germany), the manufacturer of the imaging system used in this study. F.K. is co-inventor on a patent on \ac{pa} data analysis held by FAU and iThera Medical GmbH. The remaining authors declare that they have no known competing interests.

\section*{Funding}
This research received funding from the Bavarian Ministry of Economic Affairs, Regional Development and Energy (StMWi) (DIK-2208-0014/ DIK0464/02, SENDERO) as well as from the European Research Council (ERC) under the European Union’s Horizon 2020 research and innovation programme (grant agreement No. [101002198]).

\bibliographystyle{unsrtnat}
\bibliography{references}

\end{document}

%% file: body.tex
\section{Introduction}
\label{sec:introduction}

Photoacoustic imaging (PAI) is a hybrid imaging technique that provides unique functional contrast at clinically relevant imaging depths, addressing limitations of purely optical or ultrasound-based methods. By combining light and sound, PAI visualizes the optical absorption properties of biological tissues. Short laser pulses are delivered into tissue, where absorbed radiant energy is converted into heat. This rapid heating leads to thermoelastic expansion, generating pressure waves that propagate through the tissue. These signals are then detected by \ac{us} sensors and used to reconstruct images~\cite{Beard2011-biomed-pai}.

By using laser pulses at multiple wavelengths, \ac{pai} can differentiate light-absorbing molecules (chromophores) based on their characteristic optical absorption spectra. This contrast mechanism enables the quantification of endogenous molecules such as hemoglobin, melanin, lipids, and water. As a result, \ac{pai} provides not only anatomical but also functional information, which makes it particularly well-suited for clinical imaging applications~\cite{Steinberg2019-clinical-pai}.

Numerous recent clinical studies have demonstrated the diagnostic potential of \ac{pai} across a range of diseases, particularly in oncology and vascular imaging, where its ability to distinguish between oxygenated (HbO$_2$) and deoxygenated hemoglobin (Hb) plays an important role~\cite{Knieling2025-clinical-pai}.

A fundamental challenge in \ac{pai} is fluence decay. The \ac{pa} signal is proportional not only to the tissue’s optical absorption but also to the local light fluence, which decays with depth due to scattering and absorption. This depth-dependent fluence attenuation leads to signal distortion, often resulting in the underestimation of chromophore concentrations in deeper tissue layers~\cite{Cox2012-qpat}. A related issue is spectral coloring, which occurs because the attenuation of light in tissue is wavelength-dependent. As light propagates through tissue, its intensity decreases at different rates for different wavelengths. Therefore, the measured spectrum deviates from the true absorption spectrum of the tissue, which can lead to inaccurate quantification of chromophores~\cite{rietberg2025-artifacts}.

These fluence-related effects are not just theoretical but have been confirmed by recent clinical investigations. We looked at two \ac{pai} studies on \ac{pad}, where clinicians manually placed \acp{roi} within the gastrocnemius muscle to extract oxygenation parameters~\cite{Gunther2023-pad, Caranovic2025-pad}. Analysis of these datasets revealed significant correlations between the depth of the \acp{roi} in the muscle and the measured concentrations of Hb and HbO$_2$. This depth dependency introduces a potential source of operator bias, as the apparent chromophore concentrations depend on the clinician's manual selection of the \ac{roi} position in the tissue, irrespective of the actual physiological condition. Moreover, recent clinical studies have shown the influence of skin tone on \ac{pai} measurements~\cite{Jokerst2022-ST, Else2023-ST, Wachter2026-ST}. Investigations comparing patients with varying skin pigmentation have shown that individuals with darker skin often exhibit higher estimated oxygen saturation levels in muscle tissue. Melanin, which is more abundant in darker skin, exhibits strong absorption in wavelengths that overlap with those absorbed by Hb. This superficial absorption reduces the amount of light reaching deeper tissues, disproportionately attenuating the Hb signal, leading to overestimated oxygen saturation (since oxygen saturation is calculated as the ratio between HbO$_2$ and total hemoglobin). This is consistent with observed racial bias in pulse oximetry~\cite{Sjoding2020-pulse-oxy}. 

Fluence correction is fundamentally linked to solving the optical inverse problem, which is ill-posed: an infinite number of solutions exist since the measured \ac{pa} signal depends on both the tissue’s optical absorption and the spatially varying light fluence distribution~\cite{Cox2023-optical-problem}. Accurate correction therefore requires modeling light propagation within biological tissue. Existing methods often rely on numerical approximations of the radiative transfer equation or on Monte Carlo simulations to estimate fluence distributions~\cite{Zhou2020-fc-evaluation}. While these models are capable of capturing complex light-tissue interactions, they require prior knowledge of optical tissue properties and their spatial distribution, which is rarely available in clinical settings. Current approaches often involve manual segmentation of tissue structures and the assignment of optical parameters to different tissue regions, a process that can be time-consuming for large-scale clinical datasets~\cite{Brocchu2017-fc-seg-iterative, Sweeney2025-fc-us-seg, Pattyn2021-fc-seg}. Some studies attempt to automate this process using traditional image segmentation techniques such as active contour models~\cite{Mandal2016-active-contour-fc} or graph search~\cite{Liang2022-fc-seg}, though limited tissue contrast in PA images poses challenges. Therefore, other approaches incorporate additional imaging modalities like \ac{us}~\cite{Sweeney2025-fc-us-seg}, diffuse optical tomography~\cite{Bauer2011-dot-fc}, or magnetic resonance imaging~\cite{Zhang2023-mri-fc} to provide anatomical information for fluence modeling. To reduce dependence on literature-derived optical properties, several methods employ iterative strategies to refine the tissue parameters~\cite{Cox2006-iterative-fc, Wu2021-iterative-mc, Brocchu2017-fc-seg-iterative, Zhang2022-fc-iterative}.

Recently, deep learning-based methods have been introduced as data-driven alternatives to physics-based modeling~\cite{Liang2024-fno, Madasamy2022-dl-fc, Liang2025-dl-seg-fc, Li2022-unsup-dl-fc}; these methods, however, often require access to ground truth data for training, which is rarely available for \textit{in vivo} imaging due to the inability to directly measure optical absorption or fluence distributions within tissue. As a result, most training relies on synthetic data, which rarely captures the complexity of real biological tissue. To address this, some studies have proposed the use of highly detailed anatomical digital phantoms to better bridge the simulation-to-reality gap~\cite{Lou2017-realistic-phantoms, Park2023-realistic-phantoms, gröhl2025-digitaltwins}.

In the context of skin type correction, simple empirical correction factors based on observed signal differences between skin types have been suggested~\cite{Jokerst2022-ST}. However, such approaches do not incorporate the underlying physics of light-tissue interaction and thus offer limited generalizability and accuracy.

\begin{figure}[tbp]
\centering
\includegraphics[width=\linewidth]{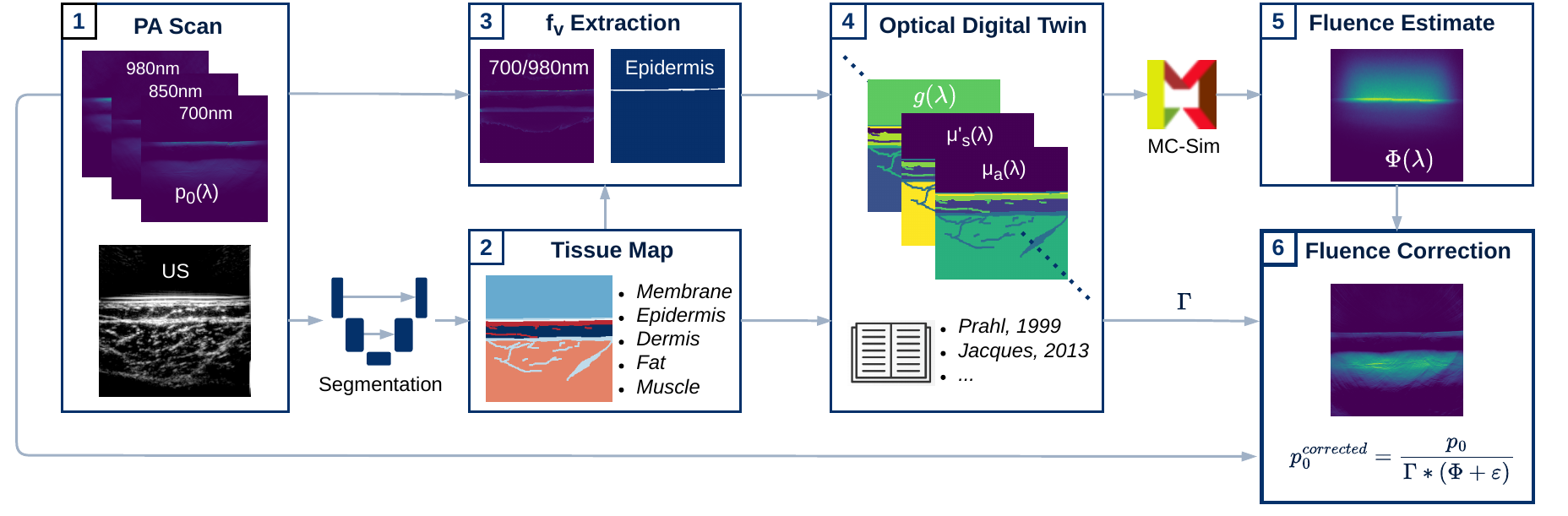}
\caption{Overview of tissue-aware fluence correction. (1) Each sample consists of multi-wavelength PA and co-registered US scans. (2) A deep neural network segments key tissue structures~\cite{Schillinger2025-pad-seg-pipeline}. (3) Epidermal melanin content $f_v$ is estimated from the 700/980\,nm signal ratio in the segmented epidermis layer. (4) Tissue maps and epidermal absorption are used to build individualized digital twins with spatially varying optical properties~\cite{Prahl-spectra, Jacques2013-tissue-params1}. (5) Monte Carlo simulations are performed on the digital twins to generate subject-specific fluence estimates $\Phi$~\cite{Grohl2022-SIMPA}. (6) The corrected image is obtained by dividing the reconstructed initial pressure $p_0$ by the simulated fluence map.}\label{overview}
\end{figure}

In this paper, we propose a subject-specific fluence correction framework that uses deep learning-based segmentation from co-registered \ac{us} images. First, a neural network classifies tissue structures, enabling the construction of a digital twin by assigning tissue-specific optical properties from literature. Monte Carlo simulations are then used to estimate the spatial distribution of light fluence, allowing for pixel-wise correction of the \ac{pa} image. To account for individual skin tone variability, melanin content is inferred from the epidermal \ac{pa} signal and incorporated into the simulation, enabling personalized, physics-informed correction. A preliminary version of this framework was presented at SPIE Photonics West 2026~\cite{Schillinger2026-fc}. To our knowledge, this is the first approach that employs deep learning–based \ac{us} segmentation to automatically generate realistic digital twins from \textit{in vivo} scans and the first to explicitly account for variations in skin color by automatically estimating melanin content directly from the acquired data (Figure \ref{overview}). 

\section{Materials and Methods}
\label{methods}

The code for our method, including implementation details, additional results, and a complete parameter list, is available on GitHub\footnote{\url{https://github.com/Mo-Sc/simpa}}.

\subsection{Ultrasound-Based Tissue Segmentation}\label{us-seg}

\Ac{pa} images often lack sufficient contrast to reliably resolve tissue boundaries. \Ac{us} images provide significantly more structural information. Since co-registered \ac{us} scans are commonly acquired alongside \ac{pa} data in many clinical \ac{pai} systems, they can be used for anatomical segmentation without introducing additional acquisition overhead.

Tissue segmentation was performed using a U-Net architecture, a convolutional neural network with an encoder-decoder structure and skip connections that is well suited for biomedical image segmentation tasks~\cite{Ronnenberger2015-unet}. Based on our previous work with similar \ac{us} datasets~\cite{Schillinger2026-optari}, we extended this architecture by incorporating an anatomy-informed geometric loss term. This additional loss penalizes anatomically implausible segmentations, such as violations in the expected vertical ordering of tissue layers (e.g., the epidermis appearing below subcutaneous fat).

The network is trained on \ac{us} images with manually annotated ground truth segmentations, as described in~\cite{Schillinger2025-pad-seg-pipeline}. The output is a multi-class segmentation map that includes the following tissue classes: membrane, \ac{us} gel, skin (subdivided into epidermis and dermis), \ac{sat}, fascia, and muscle.

\subsection{Digital Twin Construction}\label{twin-gen}

The segmented \ac{us} images were used to generate patient-specific optical property maps, creating individualized digital twins of the scanned region. For each pixel within the segmented image, we assigned wavelength-dependent optical parameters — absorption coefficient ($\mu_a$), reduced scattering coefficient ($\mu'_s$), and scattering anisotropy factor (g) — based on established values from literature. Since real tissue boundaries are not perfectly sharp, we implemented partial volume effects by mixing optical properties at segmentation borders proportionally to class probabilities.

Digital twin generation and parameter mapping were performed using the \textit{Simulation and Image Processing for Photonics and Acoustics} (SIMPA) toolkit, an open-source framework designed for simulating optical and acoustic imaging modalities~\cite{Grohl2022-SIMPA}.

The majority of the optical tissue parameters used were included in the SIMPA tissue library and are based on values compiled by Prahl~\cite{Prahl-spectra} and Jacques~\cite{Jacques2013-tissue-params1}. While these parameters are widely accepted and provide a useful approximation, they represent population averages and cannot fully capture inter-subject variability or pathological changes. This limitation is especially relevant for melanin, whose optical absorption varies strongly with skin tone. Therefore, we incorporate an explicit model for melanin-dependent absorption as described in the next section.

\subsection{Melanin Estimation and Skin Tone Modeling}

For modeling optical absorption in the skin, the relevant parameter is the volume fraction of melanosomes ($f_v$), which quantifies the percentage of tissue volume occupied by melanosomes (the organelles storing melanin). Estimates of $f_v$ in human skin range from approximately 1\% for the lightest skin types to over 40\% for the darkest~\cite{Jacques2013-tissue-params2, Vicente2022-melanin}.

Skin type is commonly categorized using scales such as the \ac{fst}. However, these scales have been widely criticized for being subjective, biased, and insufficient to capture the full range of skin tones~\cite{Okoji2021-fstlimits}. Moreover, modeling fluence based on such labels would require additional a priori information. To overcome these limitations, we estimated melanin content directly from the \ac{pa} signal in the epidermal layer. Since the epidermis is the most superficial tissue layer, its PA signal is only minimally affected by spectral coloring, so no circularity is introduced into the correction process.

In standard \ac{us} imaging, the boundary between the epidermis and dermis is difficult to distinguish. Therefore, our network only segmented a single “skin” class. To differentiate the epidermis within this class, we defined the top 100\,\textmu m (based on~\cite{Lintzeri2022-epidermis}; equivalent to 1 pixel) of the segmented skin region as the epidermal layer and the remainder as dermis. To better isolate the melanin-related contrast in the epidermis layer, we computed the ratio of \ac{pa} signal intensity at wavelengths of 700\,nm and 980\,nm. This normalization reduces the influence of patient-specific signal variability, since melanin absorption at 700\,nm is much stronger than at 980\,nm.

We translated epidermal \ac{pa} signal into $f_v$ values for simulation by defining a linear mapping such that the 5th percentile of the 700/980\,nm signal in the lightest skin type group of our test dataset (\ac{fst} I; see Section \ref{datasets}) was set to $f_v = 0.01$ and the 95th percentile in the darkest group (\ac{fst} V) was set to $f_v = 0.30$. While the true relationship between melanin content and \ac{pa} signal is likely nonlinear, this approximation provides a reasonable first-order estimate for our application.

The effective absorption coefficient of the epidermis $\mu_a^{Epi}$ at wavelength $\lambda$ is then computed as:

\begin{equation}
\mu_a^{Epi}(\lambda)=f_v \cdot \mu_a^{Mel}(\lambda) + (1-f_v) \cdot \mu_a^{Base}(\lambda)
\end{equation}

where $\mu_a^{Mel}$ denotes the intrinsic absorption coefficient of melanosomes and $\mu_a^{Base}$ the approximated baseline skin absorption, both taken from~\cite{Jacques2013-tissue-params2}.

\subsection{Modeling Light Propagation}\label{forward-model}
We estimated the spatial distribution of light fluence within tissue by running simulations using Monte Carlo eXtreme (MCX)~\cite{Fang2009-mcx} via the SIMPA framework. To model the imaging device, we adapted the preconfigured MSOT Acuity Echo scanner within SIMPA to reflect the latest hardware specifications provided by the manufacturer (iThera Medical GmbH, Munich, Germany). Because laser output power varies with wavelength in practical systems, we extended SIMPA to incorporate a spectral laser intensity profile. The wavelength-dependent laser energy values used in our simulations were obtained from the manufacturer. For each subject in all test datasets, we generated fluence maps at wavelengths of 700\,nm, 850\,nm, and 980\,nm. Each simulation was run using $10^7$ photons.

As a baseline comparison on our simulated dataset, we used a simple exponential fluence decay model based on the generalized Beer–Lambert law~\cite{Cox2023-optical-problem}:

\begin{equation}
\Phi_{exp}(z) = \Phi_0 \exp(-k z)
\end{equation}

where $z$ is the depth and $\Phi_0$ is the incident fluence taken from the simulated data. The effective attenuation parameter $k$ was chosen to best fit the dataset. The resulting exponential fluence map $\Phi_{exp}(z)$ was normalized to the average sum of the simulated fluence maps to ensure the baseline model has the same overall energy.

\subsection{Fluence Correction}

Following~\cite{Cox2023-optical-problem}, the reconstructed initial pressure distribution $p_0$ is given by the product of the local absorbed optical energy density $H$ and the \acl{pa} efficiency, which is proportional to the Grüneisen parameter $\Gamma$:

\begin{equation}
p_0(\mathbf{r}, \lambda) = H(\mathbf{r}, \lambda) \cdot  \Gamma(\mathbf{r}, \lambda)
\end{equation}

where $\mathbf{r}$ denotes the spatial position and $\lambda$ the wavelength. While $\Gamma$ depends on factors like tissue composition and wavelength, it is commonly assumed to be spatially and spectrally constant. $H$ is given as the product of the optical absorption coefficient $\mu_a$ and the local light fluence $\Phi$:

\begin{equation}
H(\mathbf{r}, \lambda) = \mu_a(\mathbf{r}, \lambda) \cdot \Phi(\mathbf{r}, \lambda)
\end{equation}

Fluence correction was thus performed by dividing the uncorrected signal $p_0$ by the simulated fluence estimate $\Phi_{sim}$ and the Grüneisen parameter $\Gamma$ at each wavelength:

\begin{equation}
p_{corr}(\mathbf{r}, \lambda)  = \frac{p_0(\mathbf{r}, \lambda)}{\Gamma \cdot (\Phi_{sim}(\mathbf{r}, \lambda) + \varepsilon)}
\end{equation}

where $p_\text{corr}$ provides a pixel-wise estimate of $\mu_a$. To ensure numerical stability in poorly illuminated regions, a small constant factor $\varepsilon$ was added and fixed across all experiments, with a value negligible relative to typical fluence values. It should be noted that the simulated fluence map $\Phi_{sim}$ itself already depends on assumed absorption coefficients taken from literature (Section \ref{twin-gen}). However, these values are used solely to estimate how light propagates through tissue, not to recover the true absorption distribution in the \textit{in vivo} scans. The resulting fluence-corrected images were spectrally unmixed using a linear unmixing approach solved via non-negative least squares at 700, 850, and 980\,nm, based on the reference absorption spectra for Hb and HbO$_2$ compiled by Prahl~\cite{Prahl-spectra}.

\subsection{Datasets and Evaluation}\label{datasets}

To validate our approach, we performed qualitative and quantitative evaluations using simulated and clinical \ac{pai} datasets:

\textbf{PAD-Dataset}. The first dataset, created in~\cite{Caranovic2025-pad}, consists of 102 human subjects, including 52 healthy volunteers and 50 patients diagnosed with intermittent claudication, a symptom of \ac{pad}. Multiple scans per subject were acquired at the gastrocnemius muscle over a 40\,mm×40\,mm field of view. The dataset includes raw \ac{pa} time series and 2D images reconstructed by the scanner using a backprojection-based algorithm. It also provides co-registered \ac{us} images and \acp{roi} manually placed in the muscle by clinical experts.

\textbf{SIM-Dataset}. To enable quantitative evaluation of our method using common image quality metrics, we created an \textit{in silico} dataset. We developed a simulation pipeline in SIMPA based on our existing framework, extended to include a full acoustic model for generating pressure time series and a backprojection-based reconstruction algorithm. As the anatomical basis for the dataset, we selected a subset of \ac{us} scans from the PAD-Dataset. For each scan, we constructed a tissue layout following the segmentation procedure described in Section \ref{us-seg}, ensuring close structural resemblance to real \textit{in vivo} scans. We extended the dataset by embedding two artificial blood vessels into the \ac{sat} layer and the gastrocnemius muscle. Vessel locations, diameters, and blood oxygenation levels were randomized within physiologically plausible ranges, mimicking structures such as the small saphenous vein. This resulted in a hybrid dataset, derived from real anatomical data but augmented with additional, controllable parameters. The dataset consists of 44 samples, simulated at 700, 850, and 980\,nm. We acknowledge that using the same forward model for both simulation and fluence correction constitutes an “inverse crime”~\cite{Kaipio2007-inversecrime}. However, this dataset primarily serves to verify the method’s functionality and quantify the contribution of individual model components. To mitigate this issue, we varied parameters and introduced random noise at multiple pipeline stages. The availability of ground truth absorption and oxygenation maps allows us to compute image quality metrics like Peak Signal-to-Noise Ratio (PSNR) and Structural Similarity Index (SSIM) for reconstructed versus true absorption images, and Mean Absolute Error (MAE) for unmixed oxygen saturation in the simulated vessels.

\textbf{FST-Dataset}. Collected in \cite{Wachter2026-ST}, this dataset includes raw \ac{pa} time series, 2D images reconstructed by the scanner using a backprojection-based algorithm, and co-registered \ac{us} scans from 46 subjects across all Fitzpatrick skin types (\ac{fst} I–VI) at different scan sites. Scans were acquired over a 40\,mm×40\,mm field of view. To ensure anatomical similarity with the segmentation training data, analysis was limited to the tibialis anterior muscle. Subjects with \ac{fst} VI were excluded due to the small sample size (n=3) and the presence of acoustic artifacts at high melanin concentrations, as reported by~\cite{Else2025-skin}. This dataset was used to investigate spectral coloring due to melanin.

\section{Results}
\label{results}

Assessing \ac{pai} performance on \textit{in vivo} data is challenging due to the lack of ground truth  and the difficulty of obtaining accurate reference measurements. Therefore, we first evaluate our method on the simulated dataset with known ground truth using common image quality metrics. For \textit{in vivo} assessment, we use the PAD-Dataset to analyze whether our approach reduces the previously discussed correlation between muscle depth and extracted PA signal intensity, whether it affects diagnostic performance, and to show representative samples before and after correction. We use the FST-Dataset to examine our method’s ability to mitigate wavelength-dependent skin-tone bias, followed by an ablation study.

\subsection{Simulated Data}\label{sim-res}

\begin{figure}[htbp]
\centering
\includegraphics[width=0.75\linewidth]{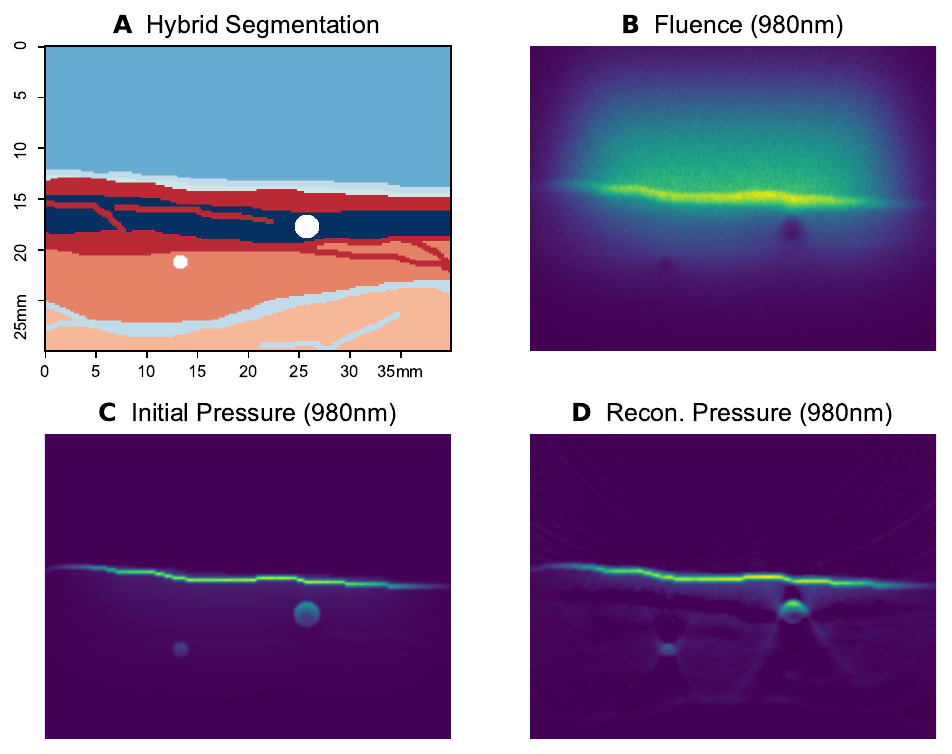}
\caption{Representative sample from the simulated dataset. \textbf{A}: Hybrid tissue mask, based on segmentation of \textit{in vivo} scans, augmented by artificial vessels. \textbf{B}: Fluence map after optical simulation at 980\,nm. \textbf{C}: Generated initial pressure distribution at 980\,nm. \textbf{D}: Reconstructed pressure after acoustic simulation and reconstruction at 980\,nm.}
\label{msot_2sim_qualresults}
\end{figure}

\begin{table}[htbp]
{
\caption{Results from the simulated dataset at 850\,nm. PSNR and SSIM at window sizes of 11 pixels (1.1\,mm) and 51 pixels (5.1\,mm) were computed for initial and reconstructed pressure images relative to ground truth absorption (higher is better). MAE of estimated oxygen saturation was evaluated in the two inserted vessels (lower is better). CI is given as 95\% confidence interval. Best values are highlighted in bold.}
\label{msot_2sim_quanresults}
\renewcommand{\arraystretch}{1.25}
\footnotesize
\begin{tabular*}{\textwidth}{@{\extracolsep{\fill}}l*{3}{c}@{}}
\toprule
Metric & \makecell{Uncorrected\\Images} & \makecell{Exponential\\Correction} & \makecell{Proposed\\Method} \\
\midrule
\multicolumn{4}{@{}l}{\textit{Initial pressure}} \\
PSNR & 19.39 {\scriptsize [19.30, 19.47]} & 20.29 {\scriptsize [20.17, 20.41]} & \textbf{27.23} {\scriptsize [26.77, 27.68]} \\
SSIM\textsubscript{11} & 0.57 {\scriptsize [0.56, 0.58]} & 0.76 {\scriptsize [0.75, 0.77]} & \textbf{0.95} {\scriptsize [0.95, 0.96]} \\
SSIM\textsubscript{51} & 0.49 {\scriptsize [0.48, 0.50]} & 0.66 {\scriptsize [0.65, 0.67]} & \textbf{0.94} {\scriptsize [0.93, 0.95]} \\
\midrule
\multicolumn{4}{@{}l}{\textit{Reconstructed pressure}} \\
PSNR & 17.89 {\scriptsize [17.73, 18.04]} & 18.88 {\scriptsize [18.81, 18.95]} & \textbf{20.57} {\scriptsize [20.45, 20.70]} \\
SSIM\textsubscript{11} & 0.52 {\scriptsize [0.49, 0.55]} & 0.58 {\scriptsize [0.57, 0.59]} & \textbf{0.60} {\scriptsize [0.59, 0.60]} \\
SSIM\textsubscript{51} & 0.52 {\scriptsize [0.50, 0.55]} & 0.58 {\scriptsize [0.57, 0.58]} & \textbf{0.68} {\scriptsize [0.67, 0.69]} \\
mSO\textsubscript{2}-MAE (Muscle-Vessel) & 11.05\% {\scriptsize [9.51\%, 12.58\%]} & 11.76\% {\scriptsize [10.20\%, 13.32\%]} & \textbf{9.58\%} {\scriptsize [7.85\%, 11.32\%]} \\
mSO\textsubscript{2}-MAE (SAT-Vessel) & 10.01\% {\scriptsize [8.91\%, 11.02\%]} & 10.59\% {\scriptsize [9.59\%, 11.56\%]} & \textbf{8.79\%} {\scriptsize [7.65\%, 9.91\%]} \\
\bottomrule
\end{tabular*}
}
\end{table}

We compared both the initial pressure and reconstructed pressure images (after acoustic modeling and backprojection-based reconstruction) to the digital twins’ ground-truth absorption values using PSNR and SSIM, computed with window sizes of 11 and 51 pixels (corresponding to 1.1\,mm and 5.1\,mm, respectively). Using multiple window sizes captures both fine structural details (small window) and larger-scale image consistency (large window). Additionally, we computed MAE between ground-truth oxygen saturation in the inserted blood vessels and reconstructed oxygenation values obtained by spectral unmixing using the three simulated wavelengths (700, 850, and 980\,nm) and dividing HbO$_2$ by total hemoglobin. Figure \ref{msot_2sim_qualresults} shows a representative digital phantom and reconstruction. Table \ref{msot_2sim_quanresults} summarizes the results at 850\,nm and compares them to the uncorrected scans as well as to the baseline exponential fluence correction model.

\subsubsection{Image Quality Metrics}

For initial pressure images, our proposed method nearly perfectly recovered the ground-truth absorption, as expected given that the forward and correction models share the same underlying simulation. The comparison with exponential correction, which performed substantially worse, highlights the benefit of accurate tissue-specific and wavelength-dependent modeling. After full acoustic modeling and reconstruction, our method still improved PSNR from 17.89 (no correction) to 20.57. SSIM increased for both small (from 0.52 to 0.60) and large (from 0.52 to 0.68) windows, although the improvement was more pronounced for the larger window, likely because fluence correction primarily removes global intensity gradients across the image, which are better captured by larger-window SSIM. Overall, the correction has a greater effect on initial pressure images, as acoustic reconstruction introduces noise and artifacts.

\subsubsection{Oxygen Saturation}

Regarding oxygenation estimates, the proposed method reduced MAE in both blood vessels by 1–2 percentage points. While the improvement is small, it should be noted that oxygen saturation is a ratio between two chromophores. Therefore, gains mainly reflect more accurate correction of spectral coloring rather than simply light attenuation with depth. In contrast, exponential correction failed to improve oxygenation estimates, likely because of its inability to model wavelength-dependent fluence effects.

\subsection{PAD-Dataset}
\subsubsection{Qualitative Examples}

\begin{figure}[tbp]
\centering
\includegraphics[width=\linewidth]{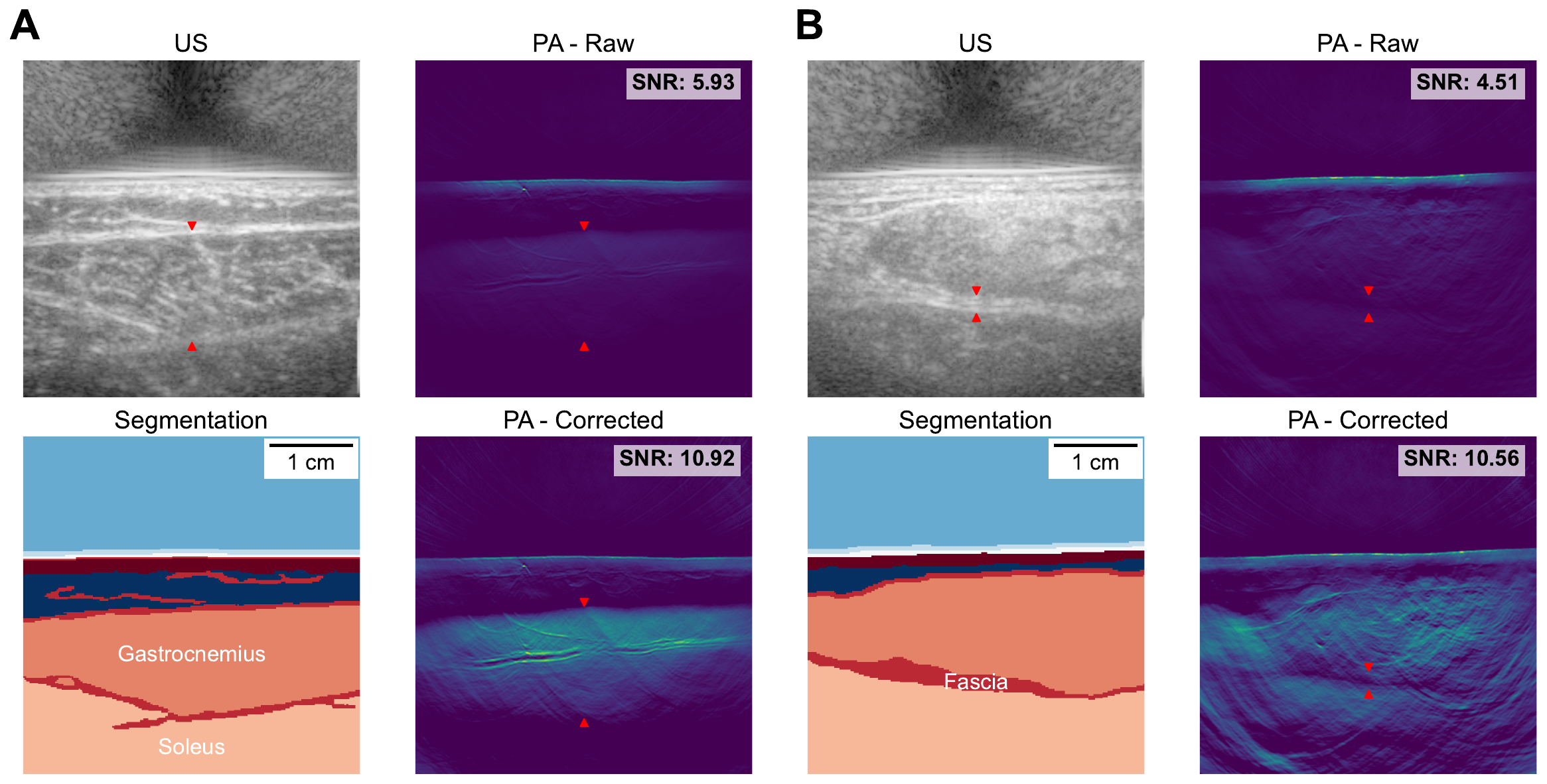}
\caption{Two representative samples from the PAD-Dataset, including \ac{us} image, uncorrected \ac{pa} image (700\,nm), the segmentation mask that was used to generate the digital twin, and the corrected \ac{pa} image. \textbf{A}: Enhanced gastrocnemius muscle. \textbf{B}: Improved visibility of the fascia, separating gastrocnemius and soleus muscle.}\label{qual_results}
\end{figure}

Figure \ref{qual_results} shows two representative samples from the PAD-Dataset before and after correction. In the first sample, the gastrocnemius muscle is barely visible in the uncorrected image but becomes clearly delineated after applying our correction method. In the second sample, the lower fascia separating the gastrocnemius and soleus muscles appears faint in the uncorrected scan but much clearer post-correction. Signal-to-noise ratio (SNR), calculated using the area outside the body as background, increased from 5.93 and 4.51 in the raw images to 10.92 and 10.56 after correction, indicating improved contrast relative to background noise. Notably, the observed structures were already present in the uncorrected images and are enhanced rather than introduced by the correction. Acoustic reconstruction artifacts remain visible and may appear more pronounced after correction.

\subsubsection{Depth–Signal Decay}

\begin{figure}[tbp]
\centering
\includegraphics[width=\linewidth]{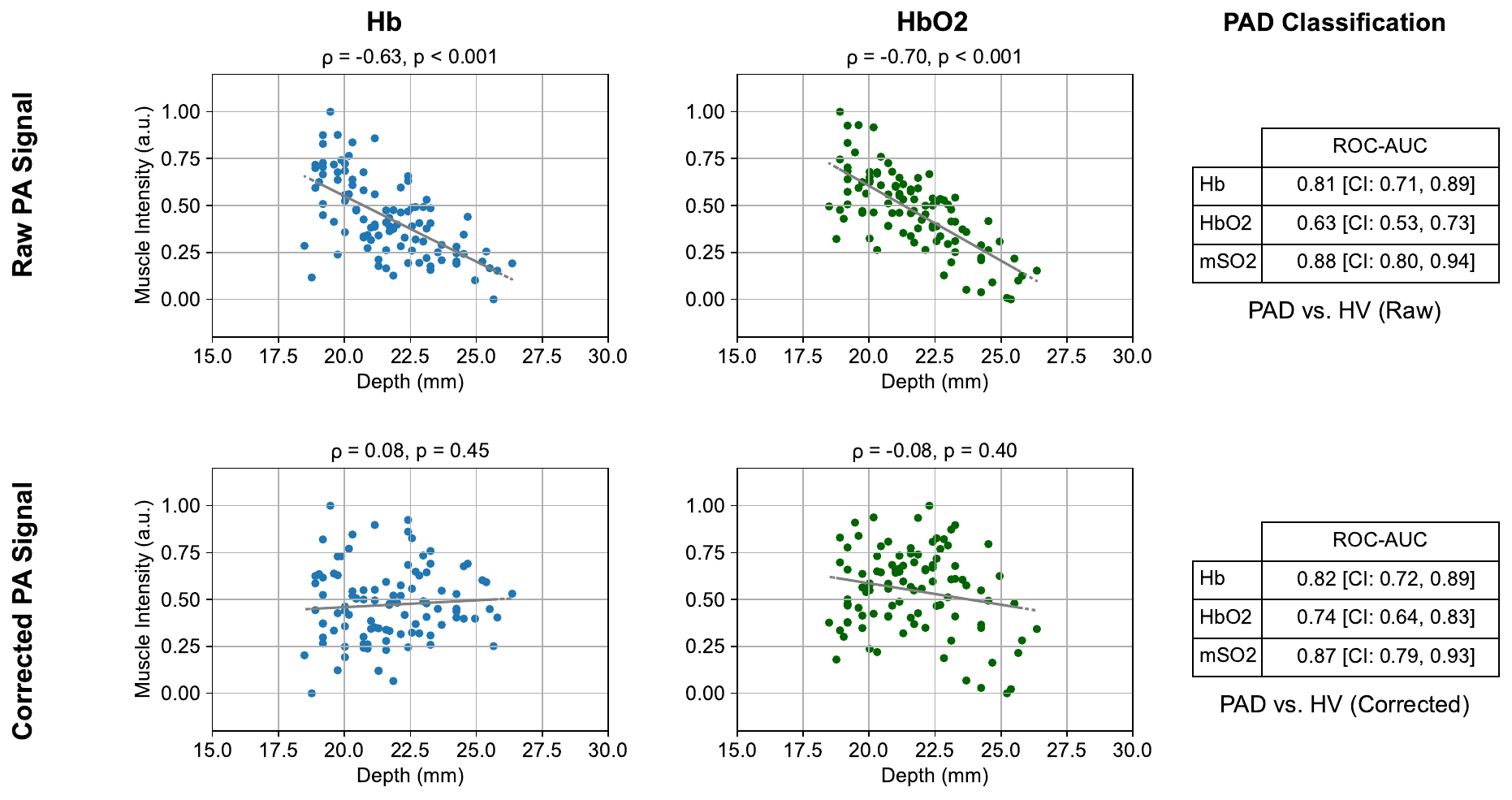}
\caption{Relationship between chromophore estimates (Hb, HbO$_2$) and the depth of the region from where they were extracted in the PAD-Dataset. Right: ROC-AUC with 95\% confidence intervals for distinguishing \ac{pad} patients from healthy volunteers (HV) based on Hb, HbO$_2$, and \ac{mso2}. Fluence correction reduces depth-related bias while maintaining or improving diagnostic performance.}\label{msot_2_results}
\end{figure}

In many clinical \ac{pai} studies, clinicians estimate chromophore concentrations inside the muscle by placing \acp{roi} into the scan and extracting signal intensity~\cite{Karlas2020-msot-pad, Knieling2017-msot-crohn, Regensburger2024-msot_ibd, Regensburger2019-msot-duchenne, Regensburger2022-msot_pediatric, Schlereth2022-msot-classification, Gunther2023-pad, Caranovic2025-pad}. Due to the depth- and wavelength dependent decay of fluence, this can introduce a systematic bias: signals from deeper \acp{roi} appear weaker, regardless of the true absorption at those locations. To investigate this effect, we analyzed the relationship between the annotated muscle depth and signal intensity in the PAD-Dataset using Spearman’s $\rho$. As shown in Figure~\ref{msot_2_results}, we observed strong negative correlations between chromophore concentration (Hb, HbO$_2$) and muscle-\ac{roi} depth before correction: $\rho = -0.63$ ($p<0.001$) for Hb and $\rho = -0.70$ ($p<0.001$) for HbO$_2$. The stronger correlation for HbO$_2$ is plausible because water absorption increases at higher wavelengths, leading to stronger attenuation for wavelengths primarily associated with HbO$_2$~\cite{Hale1973-waterspectra}. After applying our proposed method, these dependencies were substantially reduced: $\rho = 0.08$ ($p = 0.45$) for Hb and $\rho = -0.08$ ($p = 0.40$) for HbO$_2$.

\subsubsection{Diagnostic Performance}

To ensure that the correction did not degrade clinical interpretability, we re-evaluated the diagnostic performance of the data. Specifically, we replicated the classification task from the original \ac{pad} study, distinguishing patients from healthy volunteers based on photoacoustic-derived chromophore estimates~\cite{Caranovic2025-pad}. We found that diagnostic accuracy was preserved or improved after correction: the area under the receiver operating characteristic curve (ROC-AUC) for Hb and \ac{mso2} remained stable, while the AUC for HbO$_2$ increased substantially from 0.63 to 0.74.

\subsection{FST-Dataset}

\begin{figure}[tbp]
\centering
\includegraphics[width=\linewidth]{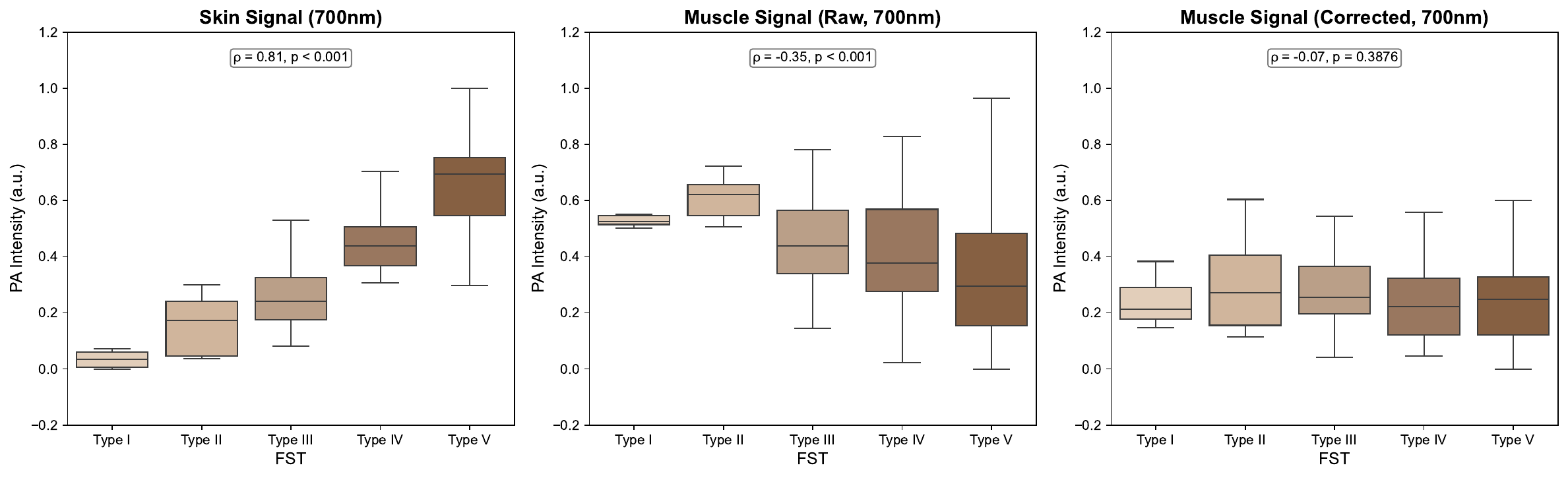}
\caption{Comparison of signal intensity at 700\,nm across Fitzpatrick skin types before and after fluence correction. Strong positive correlations were observed in the skin layer, and weaker negative correlations in the muscle layer below. Correlations with skin type were no longer significant after correction.}\label{msot3-res}
\end{figure}

To assess the proposed method’s ability to account for varying skin tones, we analyzed signal intensities across skin type groups in the FST-Dataset (Section~\ref{datasets}) at different wavelengths using Spearman's $\rho$ (Figure~\ref{msot3-res} and Table~\ref{res-ablations}). Signal intensity was first evaluated in the segmented epidermal layer. Strong positive correlations were observed between skin type and epidermal signal intensity at 700\,nm ($\rho=0.81, p<0.001$) and 850\,nm ($\rho=0.81, p<0.001$), with a weaker correlation at 980\,nm ($\rho=0.34, p<0.001$). This trend is consistent with the original publication by \cite{Wachter2026-ST} and aligns with the known absorption spectrum of melanin, which exhibits higher absorption at lower wavelengths~\cite{Jacques2013-tissue-params1}.

We then examined how skin type influences the signal intensity beneath the epidermis, specifically in the tibialis anterior muscle. As expected, stronger epidermal absorption in darker skin types resulted in reduced intensities in the muscle layer, leading to negative correlations at 700\,nm ($\rho=-0.35, p<0.001$) and 850\,nm ($\rho=-0.28, p<0.001$), with no significant correlation at 980\,nm ($\rho=-0.13, p=0.12$). Inter-patient variability was generally high, highlighting another challenge in clinical \ac{pai}.

After applying our tissue- and melanin-aware fluence correction model, correlations between skin type and muscle signals were no longer significant at any wavelength (700\,nm: $\rho=-0.07, p=0.39$; 850\,nm: $\rho=0.04, p=0.64$; 980\,nm: $\rho=-0.05, p=0.58$). These results demonstrate that personalized fluence correction maps, informed by melanin content extracted from the epidermal layer, could reduce skin-type-related bias.

\subsection{Ablation Study}

\begin{table}[tbp]
{
\caption{Results and ablation study on the \textit{in vivo} datasets. Depth and skin type correlations are reported as Spearman $\rho$ (closer to zero is better). Signal intensity was extracted from the muscle layer. Classification performance is reported as ROC-AUC with 95\% confidence interval in brackets (higher is better).}
\label{res-ablations}
\renewcommand{\arraystretch}{1.25}
\scriptsize
\setlength{\tabcolsep}{2pt}
\begin{tabular*}{\textwidth}{@{\extracolsep{\fill}}l*{6}{c}@{}}
\toprule
Metric & \makecell{No\\Correction} & \makecell{Proposed\\Method} & \makecell{Exponential\\Correction} & \makecell{Extended\\Muscle Layer} & \makecell{Extended\\SAT Layer} & \makecell{No Melanin\\Estimation}\\
\midrule
\multicolumn{7}{@{}l}{\textit{PAD-Dataset}}\\
Depth $\rho$ (Hb) & $-0.63$ (p\textless0.01) & $0.08$ (p=0.45) & $0.24$ (p=0.02) & $0.53$ (p\textless0.01) & $0.02$ (p=0.77) & $0.12$ (p=0.24)\\
Depth $\rho$ (HbO$_2$) & $-0.70$ (p\textless0.01) & $-0.08$ (p=0.40) & $-0.13$ (p=0.20) & $0.15$ (p=0.14) & $-0.14$ (p=0.16) & $-0.08$ (p=0.45)\\
ROC-AUC (\acs{mso2}) & $0.88$ {\scriptsize [0.80, 0.94]} & $0.87$ {\scriptsize [0.79, 0.93]} & $0.87$ {\scriptsize [0.80, 0.94]} & $0.87$ {\scriptsize [0.79, 0.93]} & $0.87$ {\scriptsize [0.79, 0.93]} & $0.88$ {\scriptsize [0.81, 0.94]}\\
\midrule
\multicolumn{7}{@{}l}{\textit{FST-Dataset}}\\
FST $\rho$ (700\,nm) & $-0.35$ (p\textless0.01) & $-0.07$ (p=0.39) & $-0.39$ (p\textless0.01) & $-0.05$ (p=0.55) & $-0.13$ (p=0.12) & $-0.44$ (p\textless0.01)\\
FST $\rho$ (850\,nm) & $-0.28$ (p\textless0.01) & $0.04$ (p=0.64) & $-0.35$ (p\textless0.01) & $0.03$ (p=0.69) & $-0.10$ (p=0.26) & $-0.37$ (p\textless0.01)\\
FST $\rho$ (980\,nm) & $-0.13$ (p=0.12) & $-0.05$ (p=0.58) & $-0.16$ (p=0.07) & $-0.08$ (p=0.34) & $-0.09$ (p=0.29) & $-0.18$ (p=0.03)\\
\bottomrule
\end{tabular*}
}
\end{table}

Table \ref{res-ablations} summarizes the results of the \textit{in vivo} experiments. No parameter tuning was performed for different wavelengths or datasets. In addition, we performed the following ablation studies to assess the contribution of individual model components and the robustness to segmentation inaccuracies:
\begin{itemize}
    \item \textit{Exponential Correction}: Refers to the method described in Section~\ref{forward-model}. 
    \item \textit{Extended Muscle Layer}: Proposed method, but with larger muscle layer, obtained by labeling parts of the SAT as muscle. This examines how overestimating the muscle region influences the fluence correction results.
    \item \textit{Extended SAT Layer}: Proposed method, but with larger SAT layer, obtained by labeling parts of the muscle layer as SAT. This similarly tests sensitivity to segmentation errors.
    \item \textit{No Melanin Estimation}: Proposed method without individual melanin estimation, using a constant melano\-some volume fraction across all scans.
\end{itemize}
 
\textbf{PAD-Dataset}: \textit{Exponential Correction} fails to account for wavelength-dependent factors, resulting in either undercorrection or overcorrection of Hb and HbO$_2$, regardless of the chosen decay factor $k$. For simplified tissue models, assigning too much muscle (higher absorption) results in overcorrection, likely because less light energy is assumed to reach deeper layers. The overcorrection appears stronger for Hb than for HbO$_2$, possibly because reducing the fat layer increases the relative contribution of shorter wavelengths (where Hb dominates) compared to longer wavelengths, where fat previously absorbed more strongly. Conversely, assigning too much fat (lower absorption) leads to a slightly weaker correction due to higher estimated fluence in deeper tissues. Omitting personalized melanin estimation has only a minor effect on this dataset, which consists exclusively of subjects with light skin tone. The proposed method achieves the best overall balance between Hb and HbO$_2$ correction, providing consistent improvements across both chromophores. For \ac{mso2}-based \ac{pad} classification, all methods maintained similar classification performances.

\textbf{FST-Dataset}: The negative correlation between skin type and muscle-layer signal decreased at the higher wavelengths, which is expected because melanin absorption and therefore attenuation in the skin layer decreases. Despite these wavelength-dependent differences, our proposed method corrected measurements at all three wavelengths. In contrast, \textit{Exponential Correction} and \textit{No Melanin Estimation} failed to reduce correlations with skin type, as expected due to the lack of individual melanin modeling. Simplified tissue models (\textit{Extended Muscle / SAT Layer}) that still included melanin estimation from the epidermis also reduced the skin type bias considerably, indicating that accurate modeling of melanin absorption is the dominant factor.

\section{Discussion}
\label{disc}

Our results demonstrate that a digital twin-based fluence correction model could mitigate two major sources of bias: depth-dependent attenuation and skin-tone-related signal distortion. Qualitative improvements were observed \textit{in vivo}, where anatomical structures that were previously obscured became more visible after correction. Quantitatively, we observed a substantial, wavelength-dependent reduction in the correlation between signal intensity and muscle depth, which also translated into improved PAD diagnosis based on HbO$_2$ measurements. By extracting individual melanin content from the segmented epidermal layer, we could substantially reduce wavelength-dependent intensity biases between different \acp{fst}, without prior knowledge of the subjects’ skin type. Simulation experiments confirmed improved image quality metrics and more accurate oxygenation estimates in embedded blood vessels. Ablation studies indicated that these effects result from the proposed method rather than confounding factors. Importantly, our approach is fully automated, requires no manual tissue annotation or skin type information and is therefore suitable for large-scale clinical datasets.

Apart from fluence correction, we showed how digital twins can be used to simulate realistic hybrid datasets that closely resemble \textit{in vivo} scans but can be augmented by additional structures with controllable parameters. Such digital twins could also be used for automated quality control by limiting analysis to regions where fluence is reasonably high. This would facilitate interpretation for inexperienced users, who might otherwise mistake noise for signal. Moreover, the same digital twin framework could be used to improve acoustic reconstruction, for example by providing heterogeneous speed of sound maps based on segmented tissue types, and could be further extended to incorporate additional patient metadata such as age, disease state, or other physiological factors that influence tissue composition and optical properties.

While our method accounts for several wavelength-dependent effects, some limitations remain. Discrepancies may be attributed to imperfect optical parameter spectra, reconstruction or segmentation errors, contributions from unmodeled chromophores, or dataset limitations. In particular, improvements in \ac{mso2}-based classification were limited by the strong baseline performance on the PAD-Dataset. Further validation on more challenging clinical \ac{pa} datasets is needed.

Our current segmentation model was trained on co-registered \ac{us} from specific anatomical regions and may require retraining for entirely different imaging sites. Additionally, each tissue class in the current implementation is assigned a single set of optical properties, whereas in reality these properties can vary within the same tissue type. Some parameters still have to be specified manually, and we did not evaluate the impact of stochastic variations in Monte Carlo simulations through repeated runs. The primary limitation, however, is that our simulations rely on optical tissue properties from literature. These values are based on limited experimental data and population averages and thus cannot accurately represent all patient or disease conditions. This challenge could be addressed in future implementations by incorporating iterative optimization of the optical parameters, similar to approaches proposed in~\cite{Cox2006-iterative-fc, Brocchu2017-fc-seg-iterative, Wu2021-iterative-mc}. By replacing the Monte Carlo simulation with a differentiable forward model and including acoustic simulation and reconstruction up to the reconstructed pressure maps as done in Section \ref{sim-res}, the discrepancies between simulated and measured signals could be minimized and the optical parameters refined.

In this paper, we presented a subject-specific, physics-informed framework for fluence correction in clinical \ac{pai}. Using ultrasound-guided tissue segmentation, we automatically generate individualized digital twins and estimate wavelength-dependent light fluence. Incorporating patient-specific melanin content allows for personalized correction of epidermal attenuation effects. Our work represents an important step toward robust, fair, and clinically applicable quantitative \acl{pai}.

\begin{acronym}
\acro{pai}[PAI]{photoacoustic imaging}
\acro{pa}[PA]{photoacoustic}
\acro{us}[US]{ultrasound}
\acro{sat}[SAT]{subcutaneous adipose tissue}
\acro{roi}[ROI]{region of interest}
\acrodefplural{roi}[ROIs]{regions of interest}
\acro{pad}[PAD]{peripheral artery disease}
\acro{fst}[FST]{Fitzpatrick Skin Type}
\acro{mso2}[mSO$_2$]{mean oxygen saturation}
\end{acronym}